\documentclass[12pt]{iopart}
\usepackage{graphicx}
\usepackage{cite}
\begin{document}
\title[The PCPD and non-equilibrium wetting]{The diffusive pair contact process and non-equilibrium wetting. \footnote{\it Unpublished notes intended as a basis for further research, disseminated exclusively on the cond-mat preprint server.}}
\author{Haye Hinrichsen} \vglue 3mm
\address{Theoretische Physik, Fachbereich 8, 
         Bergische Universit{\"a}t Wuppertal, \\D-42097 Wuppertal, Germany}

\begin{abstract}
The Langevin equation for the pair contact process with diffusion (PCPD) $2A\to 3A$, $2A\to\emptyset$ can be mapped by a Cole-Hopf transformation to a Kardar-Parisi-Zhang equation in a potential which has been discussed previously in the context of non-equilibrium wetting. Using this transformation the phase transition in the PCPD manifests itself as a depinning transition at the borderline of a region of phase coexistence, supporting the conjecture that the PCPD belongs to the DP universality class.
\end{abstract}

\parskip 2mm


\def\xvec{\vec{x}}			
\def\nupar{{\nu_\parallel}}		
\def\nuperp{{\nu_\perp}}		

\section{Introduction}

One of the major challenges in non-equilibrium statistical physics is the classification of phase transitions from fluctuating phases into absorbing states~\cite{Marr99,Hinr00,Odor02e}. It is believed that the critical behavior of absorbing phase transitions can be associated with a finite number of universality classes. So far only few universality classes are firmly established, the most important ones being directed percolation (DP)~\cite{Kinz83}, the parity-conserving (PC) class~\cite{GKT84,Card96}, voter-type transitions~\cite{Ligg85,Dorn01}, and the general epidemic process~\cite{Card85,Jans85}. Searching for further universality classes the pair contact process with diffusion (PCPD)
\begin{eqnarray}
\label{Process}
2A \to 3A &&  \qquad \mbox{with rate } \sigma  \nonumber\\
2A \to \emptyset &&  \qquad \mbox{with rate } \nu\\
\mbox{diffusion of individual particles} && \qquad \mbox{with rate $D$},\nonumber
\end{eqnarray}
also called annihilation-fission process, is currently one of the most promising candidates as it exhibits a continuous phase transition with an unusual type of critical behavior which has not been seen before. These exceptional properties may be related to the fact that the PCPD is a {\em binary} spreading process, i.e., {\em two} particles have meet in order to generate offspring or annihilate. 

The unusual critical behavior of binary spreading processes was first observed by Grassberger in 1982~\cite{Gras82}. The problem was then rediscovered 15 years later by Howard and T{\"a}uber~\cite{Howa97}, who proposed a bosonic field theory for the 1+1-dimensional PCPD which turned out to be unrenormalizable. More recently Carlon \etal~\cite{Carl01} investigated a `fermionic' lattice model of the PCPD model, in which the occupancy per site is restricted by an exclusion principle. Their paper trigerred a series of numerical and analytical studies~\cite{Hinr01a,Odor00,Hinr01b,Odor01,Park01,Henk01a,Odor02a,Odor02b,Noh01,Park02b,Dick02,Hinr03a,Kock02,Odor03,Bark03} and released a debate concerning the asymptotic critical behavior at the transition. Currently several viewpoints are being discussed, stating that the PCPD
\begin{enumerate}
\item represents a new universality class with a unique set of critical exponents~\cite{Hinr01a,Park02b,Kock02},
\item represents {\em two different} universality classes depending on the diffusion rate~\cite{Odor00,Odor03},
\item can be interpreted as a cyclically coupled DP and annihilation process~\cite{Hinr01b}, 
\item may be regarded as a marginally perturbed DP process with continuously varying critical exponents~\cite{Noh01},
\item may cross over to DP after very long time~\cite{Hinr03a,Bark03}.
\end{enumerate}
Each of these explanations has been supported to a different extent by physical arguments, mean field approaches, DMGR methods, and state-of-the-art simulations. The surprising variety of viewpoints demonstrates that the PCPD is a highly non-trivial process and that the resolution of these open questions is an exciting challange of non-equilibrium statistical physics.

The purpose of these notes is to point out that the Langevin equation of the PCPD is related to the problem of non-equilibrium wetting, leading to conclusions in favor of a slow crossover to DP. However, I would like to emphasize that the arguments presented here are partly speculative and need to be substantiated. Therefore these notes do not present fully validated results, rather they are intended as a basis for further research and discussions.

\section{Why DP?}

Currently most authors believe that the PCPD represents a new universality class. Depending on the model under consideration, it is observed that the asymptotic scaling regime is only reached after a long time of $10^4 \ldots 10^6$ Monte Carlo steps. The estimates for $\delta=\beta/\nupar$ seem to be close to $0.21$, while the dynamic exponenent $z\approx 1.7$ is clearly smaller than $2$, indicating superdiffusive spreading at criticality. 

\begin{figure}
\centerline{\includegraphics[width=160mm]{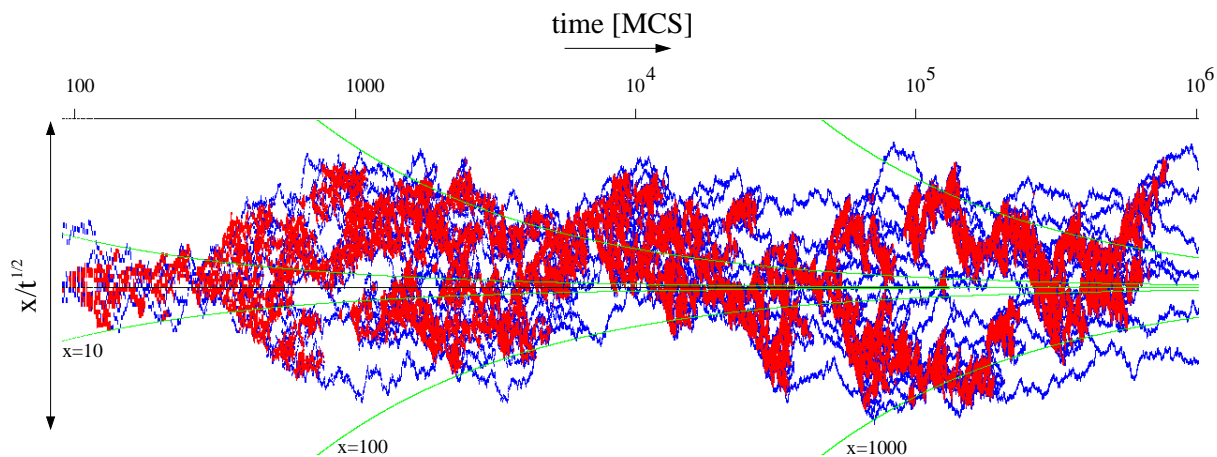}}
\caption{\label{FIGAB}
Typical spatio-temporal evolution of a binary spreading process starting
from an initial seed (figure taken from Ref.~\cite{Hinr01b}).
}
\end{figure}

This conjecture, however, poses a fundamental problem. As shown in Refs.~\cite{Hinr01b,Dick02}, binary spreading processes are characterized by two different modes (or sectors) of spreading, namely, a high-density mode dominated by self-reproducing and annihilating {\em pairs} of particles, and a low-density mode of solitary diffusing particles. The interplay of the two modes in a critical binary spreading process is illustrated in Fig.~\ref{FIGAB}, where pairs and solitary particles are represented as red and blue pixels, respectively. Plotting $x/L^{1/2}$ versus $\log_{10} t$ the figure covers four decades in time. As can be seen, patches of high activity (red) are connected by lines of diffusing solitary particles (blue). Obviously this interplay is present on all scales up to $10^6$ time steps. 

The problem arises precisely at this point: Even after $10^6$ time steps the solitary particles perform simple random walks over large distances. However, such a random walk is always characterized by the dynamic exponent $z=2$, while the process as a whole spreads superdiffusively with $z<2$. Therefore, the effective diffusion constant for solitary particles has to vary slightly under rescaling, meaning that a cluster such as in Fig.~\ref{FIGAB} cannot be scaling-invariant. Therefore it seems that the process is still far away from the asymptotic scaling regime, even after $10^6$ time steps.

Another hint can be found in the paper by Noh and Park~\cite{Noh01}, who measured the life time distribution $F(\tau)$ of solitary particles in a critical binary spreading process, finding an approximate power-law behavior $F(\tau) \sim \tau^{-\theta}$ with an exponent $\theta=2.25(5)$. Since this distribution decays faster than $\tau^{-2}$, their result would imply that the mean life time
\begin{equation}
\bar{\tau} = \frac{\int F(\tau) \, \tau \, d\tau}{\int F(\tau) \, d\tau}
\end{equation}
is finite, introducing a non-trivial time scale in the critical PCPD. The existence of such a typical life time indicates that the true asymptotic critical behavior may only be seen on extremely large scales, where $\bar{\tau}$ is virtually invisible. This scaling regime may be far beyond the accessible range of today's numerical simulations.

Observing that the numerical estimates for the critical exponents seem to move in the direction of DP values with increasing numerical effort, I suggested that a very slow crossover to DP should not be ruled out~\cite{Hinr03a}. Very recently Carlon and Barkema~\cite{Bark03} supported this point of view by a quantitative Monte Carlo and density matrix renormalization group study. In the present notes the DP hypothesis is supported in a completely different way by relating the PCPD to a non-equilibrium wetting process. However, as mentioned before, some of the arguments presented in the following are still speculative so that the hypothesis of an asymptotic DP behavior should be regarded as one out of many possible scenarios.

\section{Langevin equation for the PCPD}

In Ref.~\cite{Howa97} the Langevin equation for the ($d$+1)-dimensional PCPD was derived rigorously by introducing a bosonic operator formalism and performing the continuum limit. Using a simplified notation this Langevin equation reads
\begin{equation}
\label{Langevin}
\frac{\partial}{\partial t} \rho(\xvec,t) \;=\; b \rho^2(\xvec,t) - c \rho^3(\xvec,t) + D \nabla^2 \rho(\xvec,t) + \rho(\xvec,t)\xi(\xvec,t)\,,
\end{equation}
where $\rho(\xvec,t)$ is a coarse-grained particle density and $\xi(\xvec,t)$ denotes a white Gaussian noise with the correlations
\begin{equation}
\label{Noise}
\langle \xi(\xvec,t)\xi(\xvec',t')\rangle \;=\; 2\Gamma \,
\delta^d(\xvec'-\xvec)\delta(t'-t) \,.
\end{equation}
The four terms on the {r.h.s.} of Eq.~(\ref{Langevin}) can be interpreted as follows. Dividing the discrete lattice of the PCPD into boxes which are much larger than the lattice spacing but much smaller than the system size, $\rho(\xvec,t)$ may be understood as a coarse-grained average density of particles in a box at position $\xvec$. Assuming the particles in each box to be uncorrelated, the interplay of the binary reactions $2A \to 3A$ and $2A \to \emptyset$ leads to a quadratic term $b \rho^2(\xvec,t)$, where $b$ is essentially determined by the difference $\sigma-2\nu$ of the two reaction rates. For so-called `fermionic' models with an exclusion principle we added a cubic term $-c \rho^3(\xvec,t)$ by hand which prevents the particle density in the active phase from diverging. Moreover, there is a diffusion term and a noise field accounting for density fluctuations. 

Note that the amplitude of the noise in Eq.~(\ref{Langevin}) is proportional to the density $\rho(\xvec,t)$. This type of noise, which is known as {\em multiplicative noise} in the literature (see e.g.~\cite{Geno99}), can be motivated as follows. Since the noise accounts for fluctuations of the particle density in each box, it is primarily generated by the binary reactions $2A\to 3A$ and $2A \to \emptyset$ so that number of noise-generating sites in each box will be proportional to $\rho^2(\xvec,t)$. Thus, according to the central limit theorem, the total noise generated in the box is Gaussian and its intensity is expected to be proportional to $\rho(\xvec,t)$.

Analyzing the Langevin equation by simple power-counting one can compute the mean field critical exponents and the upper critical dimension~(see e.g.~\cite{Hinr00}). Neglecting diffusion and noise, the homogeneous stationary solution is $\rho=b/c$, hence the mean field critical point is $b_c=0$. According to the standard scaling theory of absorbing phase transitions, invariance under rescaling yields the mean-field critical exponents
\begin{equation}
\label{MFExponents}
\beta^{MF}=1,\quad 
\nuperp^{MF}=1,\quad
\nupar^{MF}=2
\end{equation}
and the upper critical dimension
\begin{equation}
d_c=2.
\end{equation}
For $d>d_c$ the coefficient $\Gamma$ scales to zero, meaning that the noise will be irrelevant on large scales so that the critical exponents are given by their mean field values~(\ref{MFExponents}). In fact, recent high-precision simulations in two spatial dimensions~\cite{Odor02a} confirm this prediction for various values of the diffusion rate. In $d<d_c$ dimensions, however, fluctuation effects lead to a non-trivial critical behavior. For this reason the present study is restricted to the (1+1)-dimensional case.

Let us first recall the main results of Ref.~\cite{Howa97}. For the unrestricted PCPD, where the cubic term is absent, the bare coefficient $b$ and the noise amplitude $\Gamma$ are related to the reaction rates in Eq.~(\ref{Process}) by
\begin{equation}
b = \sigma-2\mu \,, \qquad \Gamma = 2\sigma-\mu \,.
\end{equation}
Remarkably, in the unrestricted PCPD the critical point is always $b_c=0$, even in the presence of fluctuation effects below the upper critical dimension. This implies that the noise amplitude $\Gamma=3\sigma/4$ is positive at criticality so that the transition is characterized by `real' noise in the sense of Ref.~\cite{Howa97}. Moreover, the average particle density at the critical point was found to be constant. Regarding numerical simulations we note that this observation ubiquitously requires the Langevin equation to be iterated in the Ito sense, i.e., the density-dependent prefactor of the noise has to be evaluated {\it before} the update is carried out. Using the Stratonovich scheme one would have to introduce an additional linear drift term $-\Gamma \rho(\xvec,t)$ in Eq.~(\ref{Langevin}). 

In the inactive phase $b<0$ the annihilation process $2A\to\emptyset$ dominates so that $\rho(t)$ decays algebraically. In this case the noise amplitude is expected to become negative in the renormalization group sense, i.e., the system crosses over to `imaginary' noise after some time. For $b>0$, however, the particle density grows without limit and diverges exponentially.
        
\section{The PCPD as a non-equilibrium wetting process}

Following Ref.~\cite{Muno98} we perform a Cole-Hopf transformation
\begin{equation}
\label{ColeHopf}
h(\xvec,t) = -\ln \rho(\xvec,t)
\end{equation}
which maps Eq.~(\ref{Langevin}) to 
\begin{equation}
\frac{\partial}{\partial t} h(\xvec,t) \;=\; \Gamma - b e^{-h(\xvec,t)} + c e^{-2h(\xvec,t)}  + D \nabla^2 h(\xvec,t) - D\Bigl[\nabla h(\xvec,t)\Bigl]^2 + \xi(\xvec,t)
\end{equation}
where $\xi(\xvec,t)$ is a {\em non-multiplicative} real Gaussian noise with the same correlations as in Eq.~(\ref{Noise}). Note that the constant drift term $\Gamma$ is a consequence of the Ito interpretation in Eqs.~(\ref{Langevin})-(\ref{Noise}). Introducing the notations $\lambda=-2D$, $a=- \frac{\lambda}{2} \langle (\nabla h)^2 \rangle$, and $v_0=\Gamma-a$ this equation may be rewritten as a KPZ equation
\begin{equation}
\label{KPZ}
\frac{\partial}{\partial t} h(\xvec,t) \;=\; a - \frac{\delta V[h(\xvec,t)]}{\delta h(\xvec,t)} + D \nabla^2 h(\xvec,t) + \frac{\lambda}{2}\Bigl[\nabla h(\xvec,t)\Bigl]^2 + \xi(\xvec,t)
\end{equation}
in a potential
\begin{equation}
\label{Potential}
V[h] \;=\; \frac{ce^{-2h}}{2}  - b\,e^{-h} - v_0h\,,
\end{equation}
which has been studied recently in the context of non-equilibrium wetting~\cite{Hinr97,Hinr00b,Sant02}. Note that the constant $a$ has been defined in such a way that it compensates the average drift caused by the KPZ nonlinearity so that $v_0$ can be regarded as the average velocity of a freely evolving interface.

Let us now turn to the question how the order parameters are related in both cases.
The order parameter of the PCPD is the density of particles $\rho(\xvec,t)$. According to Eq.~(\ref{ColeHopf}) the corresponding order parameter in the wetting process  is
\begin{equation}
\rho(\xvec,t) = \exp[-h(\xvec,t)]\,.
\end{equation}
Since it is known from numerical simulations
of the PCPD that higher moments of the density $\rho^n(t)$ scale in the same way as $\rho(t)$, we may approximate the exponential function by a step function 
\begin{equation}
\exp[-h] \approx 
\left\{
\begin{array}{l}
1 \mbox { if } 0<h\leq 1 \\
0 \mbox { if } h>1
\end{array}
\right.
\end{equation}
which -- in a model with discrete heights --  is essentially the density of sites at zero height
\begin{equation}
n_0(\xvec,t) = \delta_{h(\xvec,t),0} \,.
\end{equation}
Comparing non-equilibrium wetting and the PCPD we therefore expect both quantities $\rho(\xvec,t)$ and $n_0(\xvec,t)$ to exhibit essentially the same type of asymptotic scaling behavior. Roughly speaking, the sites where the interface touches the substrate can be regarded as the active sites of the PCPD.

\section{Interpretation of the phase transition in the unrestricted PCPD}

%
%
\begin{figure}
\centerline{\includegraphics[width=100mm]{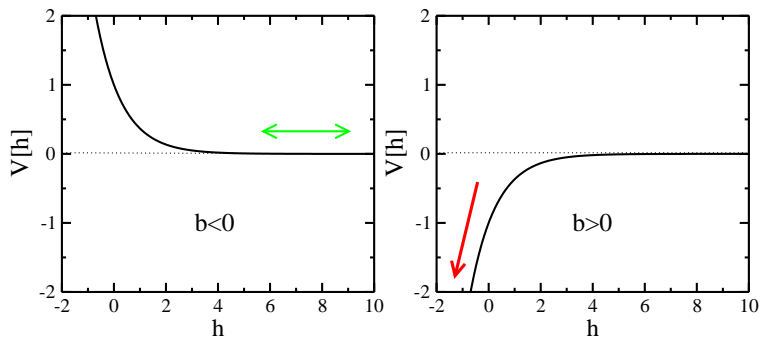}}
\caption{\label{FIGBV}
Unrestricted case:
Form of the potential $V[h]$ for $c=v_0=0$ and $b=\pm 1$. In the inactive phase $b<0$ the interface roughens close to a potential wall. In the active phase $b>0$ the interface is pulled downwards by an exponentially increasing force, corresponding to a quickly diverging particle density in the PCPD.
}
\end{figure}
In the case of the unrestricted PCPD, where multiple occupancy per site is allowed, the cubic term in Eq.~(\ref{Langevin}) vanishes so that the field theory of Ref.~\cite{Howa97} applies. As mentioned before it was shown that the transition takes place at $b=0$ even below the upper critical dimension. Moreover, the density of particles at criticality was found to be constant.

Interpreting the PCPD as a wetting process these results are easy to understand. In the inactive phase $b<0$ the potential $V[h]=-be^{-h}$ may be regarded as a lower wall representing a hard-core substrate on which the wetting layer is deposited (see Fig.~\ref{FIGFV}). As shown in~\cite{Hinr97,Muno98} the presence of a lower wall leads to a {\em continuous} wetting transition with a critical point where the propagation velocity $v_0$ of a freely evolving interface is zero. Apparently the mapping ensures that after renormalization this velocity vanishes automatically, i.e., the unrestricted PCPD is mapped {\em onto} the phase transition line of the corresponding wetting problem. Therefore, starting with a flat interface at $h=0$ (corresponding to a fully occupied lattice in the PCPD) the interface is neither pinned nor does it propagate uniformly, rather it roughens close to the wall.

In the active phase $a>0$ the potential is simply turned upside down so that an exponentially increasing force pulls the interface downwards, corresponding to a rapidly increasing particle density in the PCPD (see Fig.~\ref{FIGFV}). Therefore, in the unrestricted PCPD the transition results from a changing sign in the potential, turning the repulsive force into an attractive one, hence the transition takes place exactly at $b_c=0$. Obviously this mechanism works in any dimension and does not depend on fluctuation effects.

\section{Interpretation of the phase transition in the restricted PCPD}

Adding a cubic term with $c>0$ in the Langevin equation (\ref{Langevin}) the particle density in the active phase does no longer diverge. Such a cubic term emerges, e.g., in `fermionic' lattice models with an exclusion principle, where multiple occupancy per site is forbidden.  The cubic term can also be implemented in models with unrestricted occupancy per site by choosing the update rule in such a way that the effective fission rate decreases with increasing particle density~\cite{Kock02}. 

Although the restricted PCPD still exhibits a phase transition, its physical properties are very different:
\begin{figure}
\centerline{\includegraphics[width=60mm]{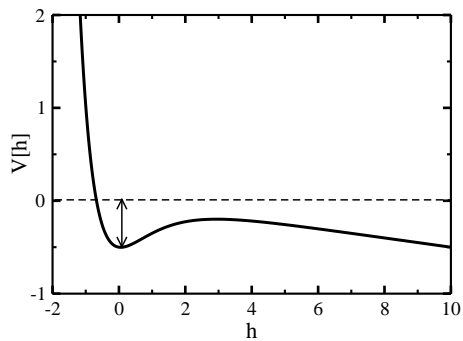}}
\caption{\label{FIGFV}
Restricted case: Form of the potential $V[h]$ for $a=b=1$. The exponential increase for $h<0$ resembles the repelling hard-core substrate while the potential well accounts for an attractive short-range force between substrate and wetting layer.  The velocity of a freely evolving interface $v_0$ is conjectured to be positive (see text), giving a slightly negative linear slope for large $h$.
}
\end{figure}
\begin{figure}
\centerline{\includegraphics[width=160mm]{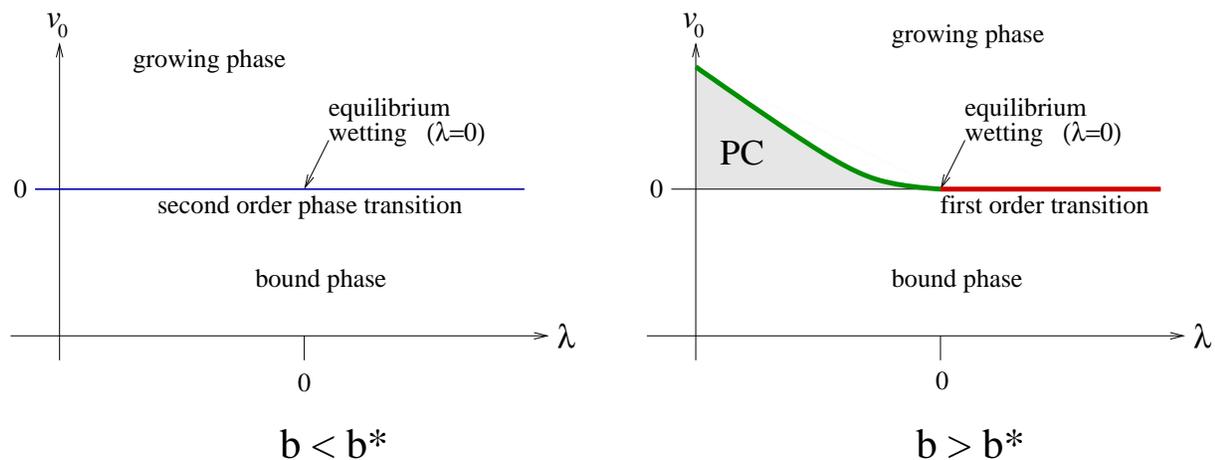}}
\caption{\label{FIGPC}
Wetting process for $c>0$. Left panel: If the attractive short range force is weak enough ($b<b^*$) the second-order wetting transition remains unaffected. Right panel: If $b>b^*$ the transition becomes first-order (red line). Moreover, for $\lambda<0$ a phase coexistence region (PC) emerges. The size of this region depends on the  value of $b$. At the upper boundary (green line) a second-order phase transition takes place. We conjecture that this transition is related to the phase transition in the PCPD.
}
\end{figure}
\begin{itemize}
\item In contrast to the unrestricted case the particle density at criticality is no longer constant, instead it decays slowly, probably as a power law with strong corrections.
\item In the active phase two different stationary states coexist, namely, the absorbing state (with $0$ or $1$ particles) and a fluctuating state with a finite density of particles.
\end{itemize}
Mapping the restricted PCPD to a wetting process we expect that the presence of a cubic term does not change the sign of the renormalized noise amplitude, i.e., right at the transition we are still dealing with `real' noise so that the Cole-Hopf transformation remains valid. As shown in Fig.~\ref{FIGFV} the cubic term gives rise to an additional potential well at zero height. This potential well may be interpreted as an attractive short-range force between substrate and wetting layer~\cite{Hinr00b,Sant02}. As a main result it was observed that such a force, if strong enough, may turn the continuous wetting transition into a discontinuous one. Moreover, in those parts of the phase diagram, where the coefficient~$\lambda$ of the KPZ nonlinearity is negative, an extended region emerges, where the bound and the moving phase coexist. The main motivation of the present notes is to relate this type of phase coexistence in the wetting process with the aforementioned coexistence of fluctuating and absorbing states in the corresponding PCPD.

The phase coexistence observed in non-equilibrium wetting works as follows. Keeping $c>0$ fixed the parameter $b$ controls essentially the depth of the potential well. If $b$ is sufficiently small the transition is not affected, i.e., it is still continuous and takes place at $v_0=0$ (see left panel of Fig.~\ref{FIGPC}). However, if the potential well is deep enough, i.e., if $b$ exceeds a certain critical threshold $b^*$, the transition becomes first order and a phase coexistence region emerges in those parts of the phase diagram where $\lambda<0$, as shown in the right panel of Fig.~\ref{FIGPC}. Within this region the short-range force is strong enough to bind the interface to the substrate although a freely evolving interface far away from the wall would already advance with the velocity $v_0>0$. 

\begin{figure}
\centerline{\includegraphics[width=150mm]{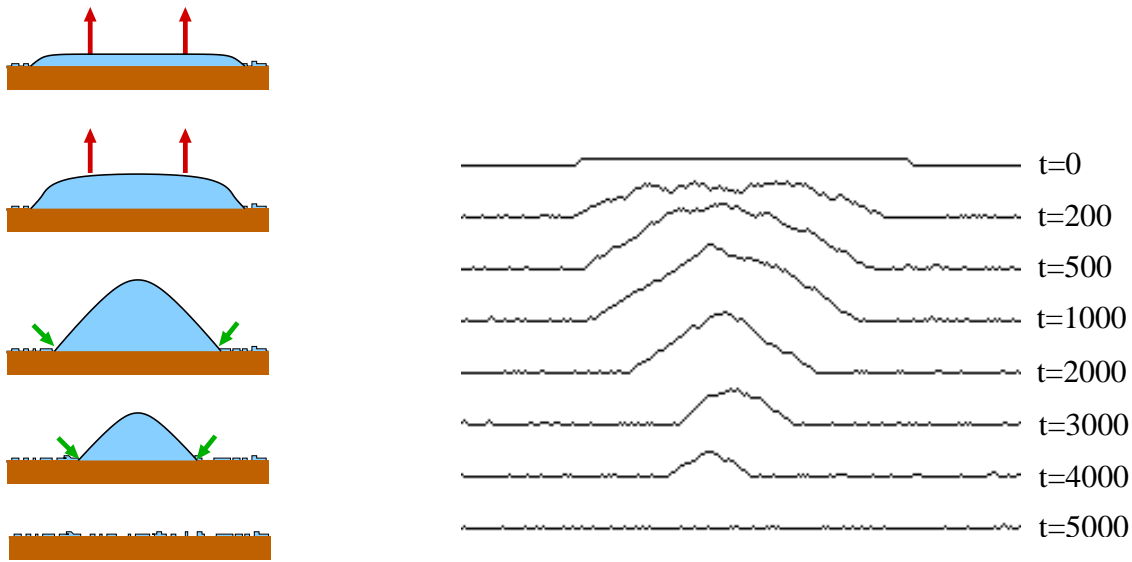}}
\caption{\label{FIGPCSTAB}
Mechanism ensuring the stability of the bound phase in the phase coexistence region. Left: If a large island is introduced by hand it first grows quickly until the edges reach a slope from where on the negative KPZ-nonlinearity suppresses further growth. Depending on $b$ the island is then "eaten up" at the outermost sites, shrinking linearly with time until it eventually disapears. Right: Corresponding simulation of a one-dimensional interface (taken from Ref.~\cite{Hinr00b}).
}
\end{figure}
The mechanism, which ensures the stability of both phases in the thermodynamic limit, can be understood as follows (see Ref.~\cite{Hinr00b}): If the interface detaches partly from the substrate due to a large fluctuation, it first advances because of $v_0>0$. The island continues to grow	 until its edges reach a certain critical slope, from where on the negative KPZ-nonlinearity suppresses further growth. The resulting pyramidial island then shrinks laterally at constant pace and eventually disapears (see Fig.~\ref{FIGPCSTAB}). The velocity at which the island shrinks is maximal at the wetting transition line $v_0=0$ and tends to zero at the upper boundary of the phase coexistence region. Obviously, this mechanism requires $\lambda$ to be negative. We note that in Eq.~(\ref{KPZ}) this is indeed the case.

Regarding the Cole-Hopf transformation we now postulate that for $c>0$ the PCPD corresponds to a wetting process for which the velocity of a freely evolving interface $v_0$ is {\em positive}. Only then the absorbing state of the PCPD, which corresponds to a completely detached interface far away from the wall, is thermodynamically stable. The postulate $v_0>0$ should emerge from a renormalization group calculation, probably due to a shift of the noise amplitude when the cubic term is introduced. It implies that the restricted PCPD corresponds to a point {\em above} the horizontal wetting transition line in Fig.~\ref{FIGPC}. Depending on the value of $b$ -- the critical parameter of the PCPD -- this point is located either inside the phase coexistence region or above. We conjecture that the coexistence region corresponds to the active phase of the PCPD and that the phase transition takes place at its upper boundary (the green line in Fig.~\ref{FIGPC}).

\section{Critical properties at the borderline of phase coexistence}

Approaching the upper boundary of the phase coexistence region the attractive short-range force becomes so weak that the velocity at which the islands shrink tends to zero. Consequently the average size of the islands in the stationary state increases, whereas the typical slope of their edges remains almost the same. 

\begin{figure}
\centerline{\includegraphics[width=140mm]{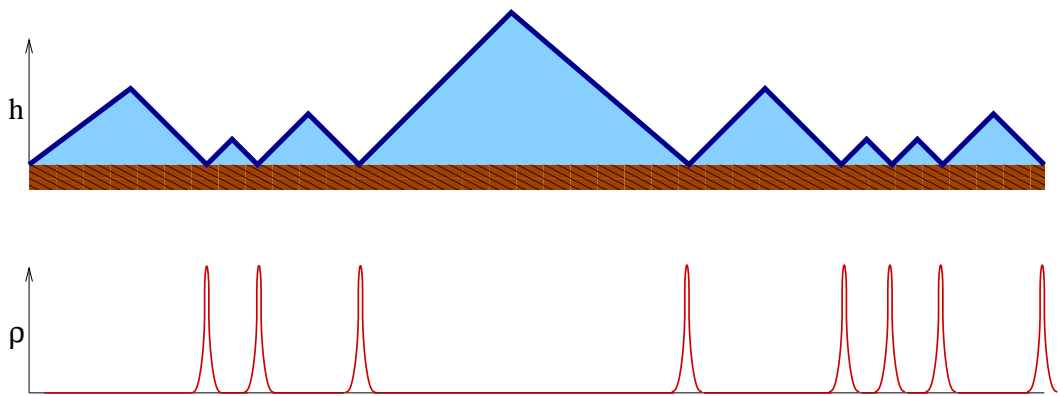}}
\caption{\label{FIGCARTOON}
Cartoon of a state in the active phase close to criticality. Top row: Typical interface profile of the wetting layer. Bottom row: Corresponding density profile $\rho=e^{-h}$ in the PCPD. The peaks represent spots of high particle density.
}
\end{figure}

A schematic illustration of a typical interface configuration in the coexistence region close to the upper boundary is shown in Fig.~\ref{FIGCARTOON}. As can be seen, the particle density $\rho=e^{-h}$ in the corresponding PCPD is indeed proportional to the density of interface sites at zero height. To understand the critical properties of the PCPD, it is therefore essential to analyze the dynamics of bottom layer sites in the corresponding wetting problem at the upper boundary of the coexistence region. Fig.~\ref{FIGSIM} shows a spatio-temporal plot of the wetting model described in Appendix A, where the bottom layer sites are represented as black pixels. If the previous assumptions are correct, these pixels should display essentially the same critical behavior as the active sites of the PCPD. Therefore, the question arises to what extent the critical dynamics of the black pixels is universal.

A possible answer is given in a recent preprint by Mu\~noz and Pastor-Satorras~\cite{Muno03}. Considering the problem of synchronization transitions in extended coupled maps they are led to exactly the same Langevin equation as in Eq.~(\ref{KPZ}). Discretizing space-time and analyzing the critical behavior at the upper boundary of the coexistence region they find numerical evidence of a directed percolation transition. This observation, together with the postulates of the present work, would imply that for $b>b^*$ the transition of the one-dimensional PCPD belongs to the DP class.

Using the language of non-equilibrium wetting the conclusion by Mu\~noz and Pastor-Satorras seems to be reasonable. The evolution of the interface is in fact dominated by the dynamics of sites at zero height. As in DP, they can spontaneously generate offspring and disappear. The large islands between those sites do not mediate effective long-range interactions, instead they seem to follow passively the dynamics at the bottom layer, adjusting their size quickly whenever their base grows or shrinks. Once the interface detaches completely, it advances with constant velocity $v_0>0$, meaning that the PCPD has entered the absorbing state.

In addition the authors of Ref.~\cite{Muno03} find a regime of first-order transitions. However, such a regime can only exist in those parts of the phase diagram where $\lambda$ is positive. In the present case, where $\lambda$ is negative, the transition belongs either to the class of multiplicative noise $(b<b^*)$ or to DP $(b>b^*)$, separated by a tricritical point at $b=b^*$ (see Ref.~\cite{Hinr00b}). Nevertheless one may observe a transient first-order behavior for $\lambda<0$, which crosses over to DP after very long time.

\begin{figure}
\centerline{\includegraphics[width=160mm]{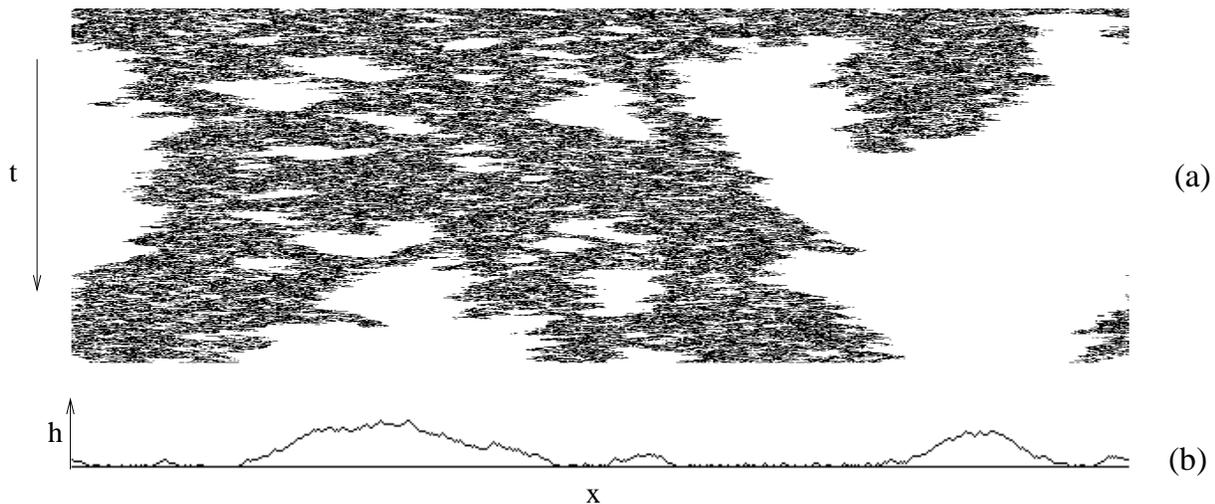}}
\caption{\label{FIGSIM}
Simualtion corresponding to the cartoon shown in Fig.~\ref{FIGCARTOON}. (a) Space-time plot of the sites at zero height (black pixels), visualizing the spatio-temporal evolution of the peaks in Fig.~\ref{FIGCARTOON}. (b) Interface configuration at the last time step of the simulation.
}
\end{figure}
%
%

\section{Conclusions}

In these notes I have propsed a relation between the diffusive pair contact process and non-equilibrium wetting, arriving at the following conclusions:

\begin{enumerate}
\item
Roughly speaking the particles in the PCPD correspond to interface sites and zero height in the corresponding wetting problem.\\

\item
The transitions in the unrestricted (bosonic) and the restricted (fermionic) PCPD rely on fundamentally different mechanisms: In the unrestricted case the transition in the corresponding wetting process is caused by a simple change of sign in the potential, whereas in the restricted case the transition emerges at the borderline of a region of phase-coexistence.\\

\item 
Referring to recent simulation results by Mu\~noz and Pastor-Satorras one can conclude that for $b>b^*$ the phase transition in the one-dimensional PCPD belongs to DP universality class, whereas for $b<b^*$ it belongs to the usual continuous wetting (multiplicative noise) universality class. Thus there may be an open door for the scenario of two universality classes depending on the models parameters.
\end{enumerate}

\noindent
Various question arise:

\begin{enumerate}
\item
{\it Where is the counterpart of the diffusive background of solitary particles?} \\
This is probably the weakest point of the continuum approach discussed here.
Obviously the continuum description does not account for single particles. Here it would be useful to study the difference between discrete and continuum models in more detail, as it has been done in the context of synchronization transitions.\\

\item
{\it Does the wetting model reproduce the algebraic decay inactive phase?}\\
No, instead one obtains an exponential decay. This failure may be related to the fact that the PCPD crosses over from `real' to `imaginary' noise when it enters the annihilation-dominated regime. By contrast the noise in the wetting model is always real.\\

\item
{\it Suppose that the PCPD belongs to the DP class, why is $d_c=2$ and not $4$ ?}\\
For $d>2$ the nonlinear term in the KPZ equation is irrelevant (unless it is very large) and thus
the phase coexistence region in the wetting problem no longer exists. Hence the DP regime cannot be accessed in $d>2$ and the model falls into the universality class of multiplicative noise.
\end{enumerate}

\noindent
Clearly the ideas presented in these notes are still speculative. Firstly, it is assumed that the Langevin description of the PCPD (including the cubic term) is valid and that the noise at criticality is Gaussian and real. Moreover, it is assumed that the Cole-Hopf transformation can be applied as usual. Finally, we postulate that the parameters renormalize in such a way that for $c>0$ the velocity of a freely evolving interface is positive. All these assumptions have to be verified. However, in my opinion the main problem of the continuum description is the missing notion of `solitary particles'. As single particles play an important role in the PCPD in restarting avalanches of high activity, it may well happen that one of the main features of the model, namely, the discrete nature of the particle density, is lost by introducing the Langevin equation. Nevertheless it is interesting to see that both problems, the PCPD and non-equilibrium wetting, are closely related and I hope that these notes may stimulate further research in this direction.

\vglue 4mm
\noindent
{\bf Acknowledgment:} I would like to thank D. Mukamel for fruitful discussions.

\vglue 4mm

\appendix
\section{A minimal model for non-equilibrium wetting}

\begin{figure}
\centerline{\includegraphics[width=90mm]{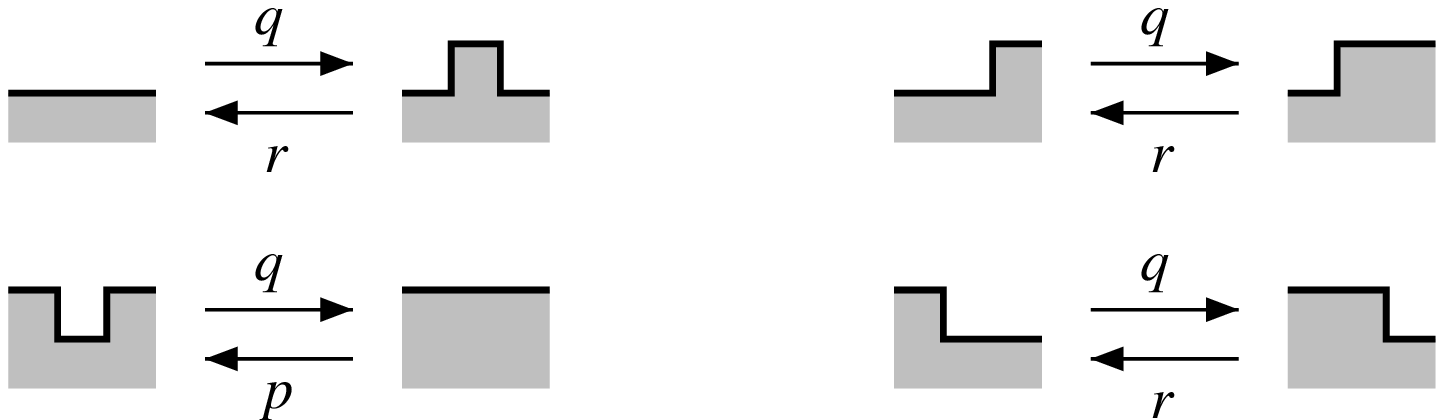}}
\caption{\footnotesize
\label{FIGRULES}
Dynamic rules of the wetting model introduced in~\cite{Hinr00b}. At the bottom layer (not shown here) evaporation is forbidden and
the deposition rate $q$ is replaced by a modified deposition
rate $q_0$ which takes the short-range interaction between substrate
and surface layer into account.
}
\end{figure}

The probably simplest model for non-equilibrium wetting, which may be regarded as a realization of Eqs.~(\ref{KPZ})-(\ref{Potential}), has been introduced some time ago in Ref.~\cite{Hinr00b}. The model is defined as a restricted solid-on-solid deposition-evaporation process of a growing interface in which the substrate is implemented as a hard-core wall at zero height. The dynamic rules involve three different elementary processes (see Fig.~\ref{FIGRULES}), namely, 
\begin{quote}
\begin{itemize}
\item[-] deposition of atoms on the substrate at rate $q_0$,
\item[-] deposition of atoms on top of islands at rate $q$,
\item[-] evaporation from the edges of islands at rate $1$, and
\item[-] evaporation from the middle of plateaus at rate $p$.
\end{itemize}
\end{quote}
The reduced growth rate $q_0$ at the bottom layer accounts for the attractive short-range force between substrate and wetting layer and determines the depth of the potential well.

\begin{figure}
\centerline{\includegraphics[width=140mm]{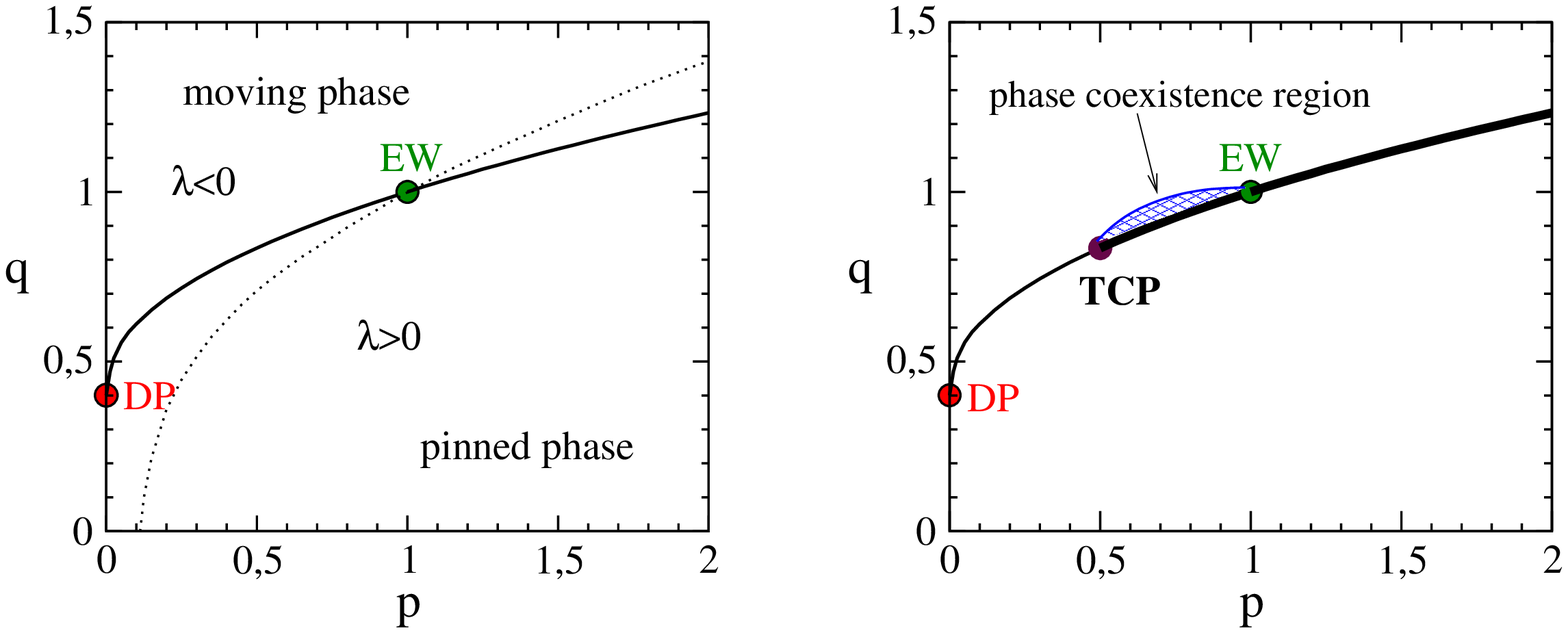}}
\caption{\footnotesize
\label{FIGPD}
Phase diagram of the wetting model introduced in~\cite{Hinr00b}. Left: Without attractive force (i.e., $q_0=q$) the wetting transition is continuous (thin line). The dotted line indicates where $\lambda$ effectively vanishes. For $p=q=1$, where both lines intersect, the dynamic rules obey detailed balance and the interface evolves according to a Edwards-Wilkinson equation in a potential (EW), while for $p=0$ the transition is determined by a directed percolation process at the bottom layer (DP). Right: Introducing a short range force by lowering $q_0$ the transition may become first order (bold line). The first order line ends in a tricritical point (TCP).
}
\end{figure}

The phase diagram of the model {\em without} attractive force $q_0=q$ is shown in the left panel of Fig.~\ref{FIGPD}. The moving phase and the bound phase are separated by a second-order transition line, where the velocity of a freely evolving interface vanishes. The line has two special points. For $p=0$ the model exhibits a special critical behavior since the transition is driven by a DP process at the bottom layer, as discussed in Refs.~\cite{Alon96,Alon98,Gold99}. Another special transition point is located at~$p=q=1$, where the dynamic rules and are symmetric under reflection $h\to-h$ so that the nonlinear term in the KPZ equation vanishes, corresponding to an Edwards-Wilkinson (EW) equation in a potential. Comparing the velocities of a horizontal and an artificially tilted interface it is possible to determine a line where the effective coefficient $\lambda$ of the nonlinear term vanishes (shown as a dotted line in Fig.~\ref{FIGPD}). As expected this line intersects the phase transition line at $p=q=1$. 

Introducing an attractive short-range force between substrate and surface layer by lowering $q_0$ the critical point $q_c(p)$ is not changed, i.e., the transition line remains the same. However, if $q_0$ is smaller than a certain threshold $q_0^*(p)$ the transition may become first order (bold line in the right panel of Fig.~\ref{FIGPD}). The first order line ends in a tricritical point (denoted as TCP) which moves along the phase transition line as $q_0$ is varied. In those parts of the phase diagram, where $p<1$ (i.e., $\lambda<0$), a phase coexistence region emerges above the transition line.

The parameter $q$ controls the growth rate and may be associated with $v_0$, while $q_0$ determines the strength of the short range force which is related to the parameter $b$ in the Langevin equation. The parameter $p$ can be used to control the effective value of $\lambda$, the coefficient of the nonlinear term in the KPZ equation.

\newpage
  

\bibliographystyle{paper}	
{\noindent \large \bf References:}
\bibliography{paper}		

\begin{thebibliography}{10}

\bibitem{Marr99}
Marro J and Dickman R.
\newblock {\em Nonequilibrium phase transitions in lattice models}.
\newblock Cambridge University Press, Cambridge, (1999).

\bibitem{Hinr00}
Hinrichsen H, {\em Nonequilibrium critical phenomena and phase transitions into
  absorbing states}, {Adv. Phys.} 49, 815, also available as eprint
  cond--mat/0001070 (2000).

\bibitem{Odor02e}
{\'O}dor G.
\newblock Phase transition universality classes of classical, nonequilibrium
  systems.
\newblock unpublished, eprint cond-mat/0205644, (2002).

\bibitem{Kinz83}
Wolfgang Kinzel.
\newblock Percolation structures and processes.
\newblock In G.~Deutscher, R.~Zallen, and J.~Adler, editors, {\em Ann. Isr.
  Phys. Soc.}, volume~5, Bristol, (1983). Adam Hilger.

\bibitem{GKT84}
Grassberger P, Krause F, and von~der Twer~T, {\em A new type of kinetic
  critical phenomenon}, {J. Phys. A} 17, L105 (1984).

\bibitem{Card96}
Cardy J and T{\"a}uber~U C, {\em Theory of branching and annihilating random
  walks}, {Phys. Rev. Lett.} 77, 4780 (1996).

\bibitem{Ligg85}
Liggett~T M.
\newblock {\em Interacting particle systems}.
\newblock Springer, Berlin, (1985).

\bibitem{Dorn01}
Dornic I, Chat{\'e} H, Chave J, and Hinrichsen H, {\em Critical coarsening
  without surface tension: the voter universality class}, {Phys. Rev. Lett.}
  87, 5701 (2001).

\bibitem{Card85}
Cardy~J L and Grassberger P, {\em Epidemic models and percolation}, {J. Phys.
  A} 18, L267 (1985).

\bibitem{Jans85}
Janssen~H K, {\em Renormalized field theory of dynamical percolation}, {Z.
  Phys. B} 58, 311 (1985).

\bibitem{Gras82}
Grassberger P, {\em On phase transitions in schl{\"o}gl's second model}, {Z.
  Phys. B} 47, 365 (1982).

\bibitem{Howa97}
Howard~M J and T{\"a}uber~U C, {\em `real' versus `imaginary' noise in
  diffusion-limited reactions}, {J. Phys. A} 30, 7721 (1997).

\bibitem{Carl01}
Carlon E, Henkel M, and Schollw{\"o}ck U, {\em Critical properties of the
  reaction-diffusion model $2a \to 3a$, $2a \to \emptyset$}, {Phys. Rev. E} 63,
  036101 (2001).

\bibitem{Hinr01a}
Hinrichsen H, {\em Pair contact process with diffusion: A new type of
  nonequilibrium critical behavior?}, {Phys. Rev. E} 63, 036102 (2001).

\bibitem{Odor00}
{\'O}dor G, {\em Critical behavior of the one-dimensional annihilation-fission
  process $2a \to \emptyset$, $2a \to 3a$}, {Phys. Rev. E} 62, R3027 (2000).

\bibitem{Hinr01b}
Hinrichsen H, {\em Cyclically coupled spreading and pair annihilation},
  {Physica A} 291, 275 (2001).

\bibitem{Odor01}
{\'O}dor G, {\em Phase transition of the one-dimensional coagulation-production
  process}, {Phys. Rev. E} 63, 067104 (2001).

\bibitem{Park01}
Park K, Hinrichsen H, and Kim I-M, {\em Binary spreading process with parity
  conservation}, {Phys. Rev. E} 63, R065103 (2001).

\bibitem{Henk01a}
Henkel M and Hinrichsen H, {\em Exact solution of a reaction-diffusion process
  with three-site interactions}, {J. Phys. A} 34, 1561 (2001).

\bibitem{Odor02a}
{\'O}dor G, Marques~M C, and Santos~M A, {\em Phase transition of a
  two-dimensional binary spreading model}, {Phys. Rev. E} 65, 056113 (2002).

\bibitem{Odor02b}
{\'O}dor G, {\em Multicomponent binary spreading process}, {Phys. Rev. E} 66,
  026121 (2002).

\bibitem{Noh01}
Noh~J D and Park H.
\newblock Novel universality class of absorbing transitions with continuously
  varying exponents.
\newblock unpublished, eprint cond-mat/0109516, (2001).

\bibitem{Park02b}
Park K and Kim I-M, {\em Well-defined set of exponents for a pair contact
  process with diffusion}, {Phys. Rev. E} 66, 027106 (2002).

\bibitem{Dick02}
Dickman R and de~Menezes M A~F, {\em Nonuniversality in the pair contact
  process with diffusion}, {Phys. Rev. E} 66, 045101 (2002).

\bibitem{Hinr03a}
Hinrichsen H, {\em Stochastic cellular automaton for the coagulation-fission
  process $2a \to 3a$, $2a \to a$}, {Physica A} 320, 249 (2003).

\bibitem{Kock02}
Kockelkoren J and Chat{\'e} H.
\newblock Absorbing phase transitions of branching annihilating random walks.
\newblock unpublished, eprint cond-mat/0208497, (2002).

\bibitem{Odor03}
{\'O}dor G, {\em On the critical behavior of the one-dimensional diffusive pair
  contact process}, {Phys. Rev. E} 67, 016111 (2003).

\bibitem{Bark03}
Barkema~G T and Carlon E.
\newblock Universality in the pair contact process with diffusion.
\newblock eprint cond-mat/0302151, (2003).

\bibitem{Geno99}
Genovese W. and Mu{\~n}oz~M A, {\em Recent results on multiplicative noise},
  {Phys. Rev. E} 60, 69 (1999).

\bibitem{Muno98}
Mu{\~n}oz~M A and Hwa T, {\em On nonlinear diffusion with multiplicative
  noise}, {Europhys. Lett.} 41, 147 (1998).

\bibitem{Hinr97}
Hinrichsen H, Livi R, Mukamel D, and Politi A, {\em A model for nonequilibrium
  wetting transitions in two dimensions}, {Phys. Rev. Lett.} 79, 2710 (1997).

\bibitem{Hinr00b}
Hinrichsen H, Livi R, Mukamel D, and Politi A, {\em First order phase
  transition in a 1+1-dimensional nonequilibrium wetting process}, {Phys. Rev.
  E} 61, R1032 (2000).

\bibitem{Sant02}
de~los Santos~F, Telo da~Gama M~M, and M{\~u}noz~M A, {\em Stochastic theory of
  non-equilibrium wetting}, {Europhys. Lett.} 57, 803 (2002).

\bibitem{Muno03}
Munoz~M A and Pastor-Satorras R.
\newblock Stochastic theory of synchronization transitions in extended systems.
\newblock eprint cond-mat/0301059, (2003).

\bibitem{Alon96}
Alon U, Evans~M R, Hinrichsen H, and Mukamel D, {\em Roughening transition in a
  one-dimensional growth process}, {Phys. Rev. Lett.} 76, 2746 (1998).

\bibitem{Alon98}
Alon U, Evans~M R, Hinrichsen H, and Mukamel D, {\em Smooth phases, roughening
  transitions and novel exponents in one-dimensional growth models}, {Phys.
  Rev. E} 57, 4997 (1998).

\bibitem{Gold99}
Goldschmidt~Y Y, Hinrichsen H, Howard~M J, and T{\"a}uber~U C, {\em Novel
  nonequilibrium critical behavior induced by unidirectional coupling of
  stochastic processes}, {Phys. Rev. E} 59, 6381 (1999).

\end{thebibliography}
\end{document}